\documentclass[authoryear,final,3p,times]{elsarticle}

\usepackage[colorlinks=true,linkcolor=blue]{hyperref}

\usepackage{amssymb}

\usepackage{siunitx} 
\usepackage{amsmath} 
\usepackage{amsfonts}
\usepackage{amsxtra}
\usepackage{color}
\newcommand{\reff}[1]{Fig. \ref{#1}} 
\newcommand{\reft}[1]{Tab. \ref{#1}} 
\newcommand{\refe}[1]{(\ref{#1})}    

\newcommand\matr[1]{\underline{\underline{\mathbf{#1}}}}
\newcommand\vect[1]{\underline{\boldsymbol{#1}}}

\usepackage{ulem}

\journal{{\normalfont Accepted manuscript, Journal of Materials Processing Technology, \urlstyle{same} \url{https://doi.org/10.1016/j.jmatprotec.2016.10.010}}}

\usepackage{etoolbox}
\makeatletter
\patchcmd{\ps@pprintTitle}% <cmd>
  {Preprint submitted to}% <search>
  {}% <replace>
  {}{}% <succes><failure>
\makeatother

\newcommand\blfootnote[1]{%
  \begingroup
  \renewcommand\thefootnote{}\footnote{#1}%
  \addtocounter{footnote}{-1}%
  \endgroup
}

\begin{document}

\begin{frontmatter}


\title{Simulation of self-piercing rivetting processes in fibre reinforced polymers: material modelling and parameter identification}

\author[label1]{Franz Hirsch}
\author[label1]{Sebastian M\"uller}
\author[label2]{Michael Machens}
\author[label3]{Robert Staschko}
\author[label2]{Normen Fuchs}
\author[label1,label4]{Markus K\"astner\corref{cor1}}\ead{markus.kaestner@tu-dresden.de}
\address[label1]{TU Dresden, Institute of Solid Mechanics, D-01062 Dresden}
\address[label2]{Fraunhofer Application Center Large Structures in Production Technology AGP, D-18059 Rostock}
\address[label3]{University of Rostock, Chair of Manufacturing Engineering, D-18059 Rostock}
\address[label4]{TU Dresden, Dresden Center for Computational Materials Science, D-01062 Dresden}
\cortext[cor1]{Corresponding author, +49 351 463 32656}

\blfootnote{© 2017. This manuscript version is made available under the CC-BY-NC-ND 4.0 license}
\blfootnote{https://creativecommons.org/licenses/by-nc-nd/4.0/}

\begin{abstract}
This paper addresses the numerical simulation of self-piercing rivetting processes to join fibre reinforced polymers and sheet metals. Special emphasis is placed on the modelling of the deformation and failure behaviour of the composite material. Different from the simulation of rivetting processes in metals, which requires the modelling of large plastic deformations, the mechanical response of composites is typically governed by intra- and interlaminar damage phenomena. Depending on the polymeric matrix, viscoelastic effects can interfere particularly with the long-term behaviour of the joint. We propose a systematic approach to the modelling of composite laminates, discuss limitations of the used model, and present details of parameter identification.
 
Homogenisation techniques are applied to predict the mechanical behaviour of the composite in terms of effective anisotropic elastic and viscoelastic material properties. In combination with a continuum damage approach this model represents the deformation and failure behaviour of individual laminae. Cohesive zone elements enable the modelling of delamination processes. The parameters of the latter models are identified from experiments. The defined material model for the composite is eventually utilised in the simulation of an exemplary self-piercing rivetting process.
\end{abstract}

\begin{keyword}
Self-piercing rivetting \sep Fibre-reinforced composites\sep Damage \sep Homogenisation \sep Cohesive Zone
\end{keyword}

\end{frontmatter}

\section{Introduction}
\label{sec:introduction}

In the light of an increasing ecological awareness, the consistent lightweight design has become a major focal point of research and development especially in the automotive, railway and aircraft industries. Hybrid designs that combine different materials, e.g. metals and fibre reinforced polymers (FRP), are considered a promising approach to cost-effective lightweight components. While these multi-material designs provide considerable synergy effects by merging advantageous properties of various materials, joining technologies are essential for the practical application of hybrid materials and structures. 

Self-Piercing Rivetting (SPR) has been demonstrated to be a fast and robust technique to join dissimilar sheet metals, see for instance the review of \citet{He2008}. In contrast to bolted or rivetted joints, there is no need for pre-drilled holes in the SPR process which reduces production costs and time significantly. In addition, process control is well established for SPR which ensures the reproducibility of joint properties. 

\begin{figure}
  \begin{center}
    (a)~\includegraphics[scale=0.8]{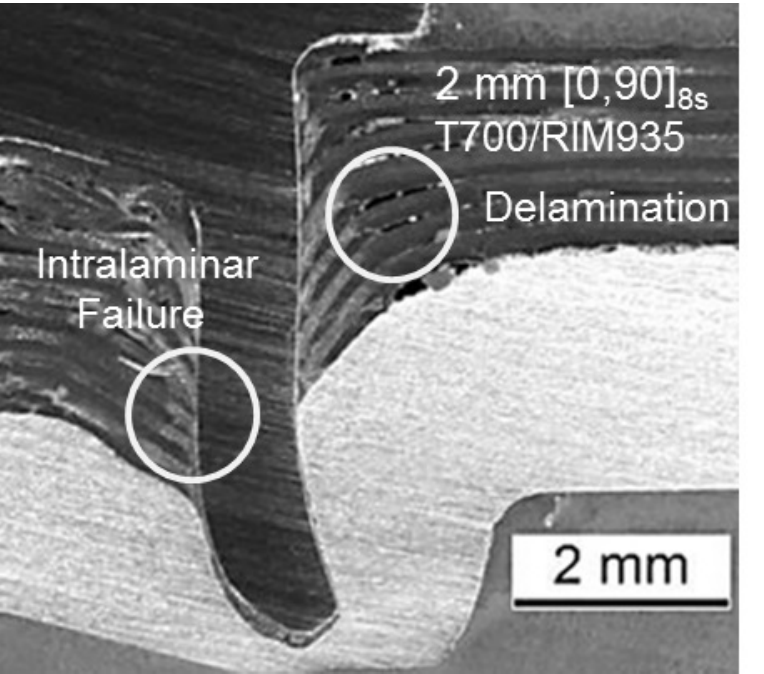}\qquad
    (b)~\includegraphics[scale=0.8]{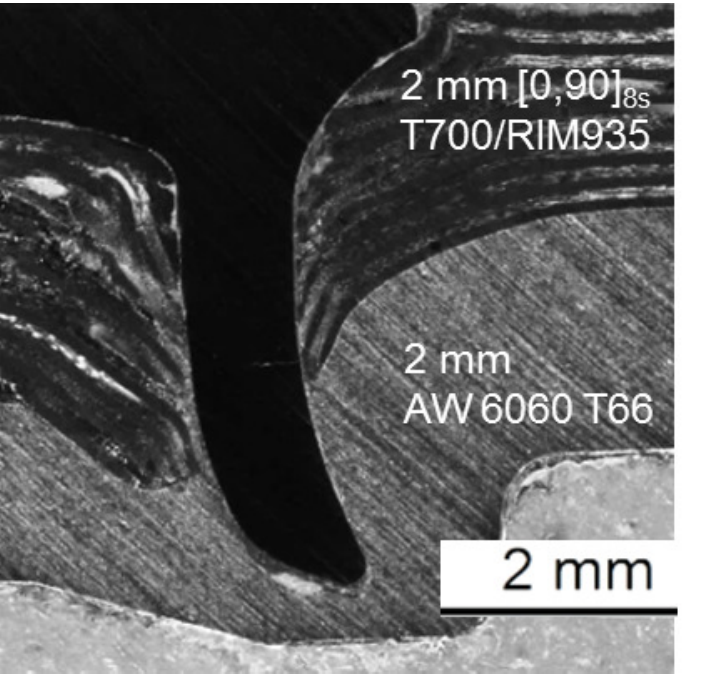}
    \caption{Micrograph images of two SPR joints show severe damage of the FRP in the vicinity of the rivet: (a) flat head rivet, and (b) countersunk head rivet.
      With interlaminar damage due to fibre and interfibre cracks, and interlaminar damage, i.e. delamination, two principal phenomena of the
      damage behaviour of FRP can be identified.}
  \label{fig:RivetDamage}
  \end{center}
\end{figure}

\cite{Fratini2009} have presented the first application of the SPR technology to join FRP and metals. Different from the ductile deformation behaviour of metals, FRP materials generally exhibit a limited strain to failure, especially if epoxy resins are used as matrix materials. During the piercing of the FRP material, reinforcing fibres are cut and the complex stress state in the process zone induces further damage, e.g. matrix cracks or failure of the fibre-matrix interface. In addition to these intralaminar damage and failure phenomena, laminated composite materials are prone to interlaminar failure. These delaminations are caused by the bending deformation of the laminate and compressive in-plane stresses that result from the expansion of the rivet, \reff{fig:RivetDamage}.

The described damage phenomena in the composite determine the setting force characteristics in the first stage of the SPR process, and the induced damage can influence the strength of the joint. 
Therefore, the effect of selected process, design, and material parameters on the damage in the composite as well as on the strength of the joint has already been studied experimentally.
Regarding the composition of the FRP material, thermoset and thermoplastic matrices have been combined with carbon and glass fibres.
In the context of carbon fibre reinforced thermosets, the group of Di Franco et al. conducted static and fatigue test with single lap joint specimens. 
In \cite{DiFranco2012} they consider the influence of the distance between two rivets in hybrid SPR joints in combination with adhesive bonding. 
To find optimized joint configurations, further studies on several configurations were made in \cite{DiFranco2013a}. In addition to geometric parameters of the specimens, 
the influence of process parameters were investigated in \cite{DiFranco2013}. Similar studies of process parameters and specimen configurations for glass fibre reinforced 
thermosets were performed by \cite{Fratini2009}. They investigated three different cases with varying thickness of the composite and aluminium layers.
In the field of reinforced thermoplastics \cite{Zhang2014} investigate SPR parameters of glass and carbon reinforced PA6. They evaluated strength and failure mechanisms for single lap joint tests.
A major influence is expected from the shape of the rivet. While semi-tubular rivets with different heads have been used in the studies mentioned above, solid punch rivets were 
considered by \cite{Meschut2014}. In addition to mechanical loads also thermal loads influence the strength of SPR joints. Their effect on the damage propagation in carbon/glass fibre hybrid 
thermosets was investigated by \cite{Wagner2014} where thermal loads are found to cause a reduction of the joint strength.  

The robust simulation of mechanical joining processes poses a challenge due to the involved non-linearities, e.g. large plastic deformations, damage and failure of the joined materials, 
and multi-body contact, that have to be accounted. For the simulation of SPR processes based on the Finite Element Method (FEM) algorithms for the deletion of distorted elements based on 
various geometric and physical criteria have been developed. A 2D rivetting process with two aluminium layers was simulated by \cite{Atzeni2009}. Cutting of the top layer was modelled in terms 
of a continuum damage model in combination with element deletion. A comparison of the deformed geometry and the load-displacement paths shows good agreement 
between experiment and simulation. \cite{Casalino2008} investigated effects of the mesh size in the penetration zone of the top layer. They used a similar set up with a ductile fracture model 
and element deletion. The force-displacement curves and the deformed shape were compared to experiments and a good agreement was found for sufficiently refined meshes. 
\cite{Bouchard2008} performed simulations of SPR processes with different geometric configurations and material properties showing the robustness and applicability of 
the FE simulation of SPR processes with two metal layers. However, the mentioned modelling approaches based on the deletion of elements which may lead to an artificial 
mass loss due to the removed elements. Moreover, fine meshes are required to obtain smoothly eroded interfaces. \cite{Porcaro2006b} resolve this problems with the help of an adaptive 
mesh algorithm where the nodal positions are relocated to improve the element shapes within a defined time interval. These two major approaches are available in commercial FEM tools and 
have been successfully applied to the simulation of SPR processes for joining various metals, see the comprehensive review by \cite{He2012}. Consequently, the numerical simulation can 
now be considered a reliable tool for the development, analysis, and evaluation of mechanical joining techniques as far as metals are considered \citep{Eckstein2007}. Simulations that 
involve composites require further research and verified procedures have to be established.

\begin{figure}
 \begin{center}
  \includegraphics[scale=1]{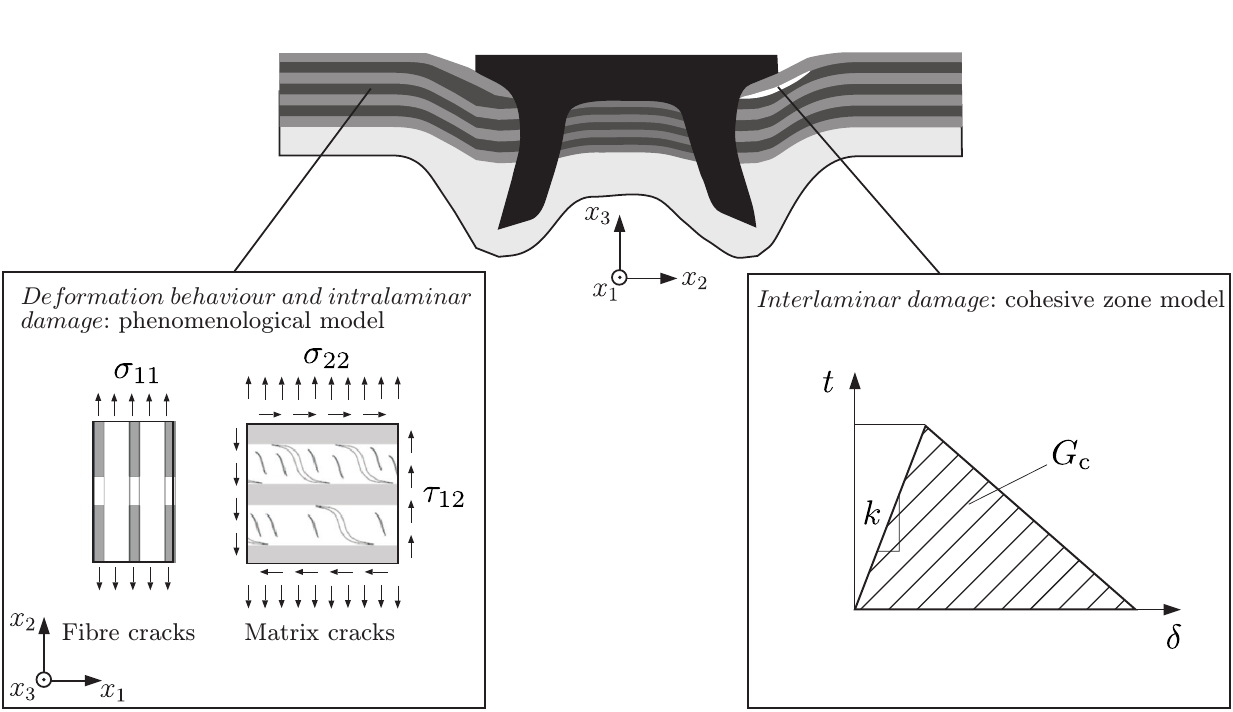}
  \caption{Numerical modelling of layered FRP materials in the simulation of SPR processes. Intralaminar damage, e.g. due to fibre and matrix cracks, is accounted in terms of a phenomenological model while delamination is modelled by a cohesive zone model.}
  \label{fig:Methodical_Framework}
 \end{center}
\end{figure}

This contribution therefore addresses the FE simulation of SPR processes for FRP materials. Beyond a state of the art modelling of the rivet and the metal bottom sheet, reasonable predictions for the punch force and for the damage induced in the composite sheet require primarily an accurate numerical model of the deformation, damage, and failure behaviour of the composite. Here, the FRP material is assumed to have a laminated structure and the following approach is applied to model the laminate, \reff{fig:Methodical_Framework}:
\begin{enumerate}
  \item \emph{Deformation behaviour and intralaminar damage/failure:} The an\-i\-so\-tro\-pic mechanical behaviour of the individual composite layer is modelled in terms of a phenomenological material model that has to account for the undamaged deformation behaviour as well as for fibre failure and matrix damage. The failure of the layer is represented by the deletion of elements.
  \item \emph{Interlaminar damage/failure:} In between elements that represent two different layers, contact conditions are included that model delamination in terms of a cohesive zone model.
\end{enumerate}

A similar modelling strategy has been used by \cite{Gruetzner2013} to qualitatively analyse the damage in the composite induced 
during the piercing of the laminate. In addition, an SPR process has been simulated but no quantitative comparison of the load-displacement curves nor details on the used material parameters have been presented. \cite{Kroll2013} have simulated the SPR process 
and analysed the strength of the joint for the combination of quasi-isotropic FRP laminates and aluminium. As their modelling approach does not account for the layered structure of the composite, 
it is not possible to study the influence of delamination and intralaminar damage on the quality of the joint.

The choice of the material models that represent the individual layers of the composite depends on the material behaviour and the purpose of the simulation. If the degradation of the material 
is negligible, failure criteria for composites which are available in commercial FE software can be used. For different failure modes observed in composites \cite{Hashin1980} developed 
three-dimensional failure criteria in terms of quadratic stress polynomials. \cite{Puck1998} introduced new criteria for inter-fibre failure
to account for brittle failure of composites. The failure criteria are evaluated in terms of stress acting on characteristic action planes associated to inter-fibre failure.
With the failure mode concept, \cite{Cuntze2004} presented a method to model failure of composites as a combination of nine individual failure modes.

If the gradual degradation of the material in the process zone is of interest, e.g. to evaluate the resulting strength of the joint, damage models based 
on the concept of continuum damage mechanics \citep{Lemaitre1996} can be used to represent the individual laminates. For isotropic damage phenomena simple 
models with a single scalar variable are common approaches. Anisotropic formulations that distinguish characteristic failure modes, e.g. matrix and interface 
degradation as well as fibre failure, are available in the literature. For the two-dimensional case \cite{Matzenmiller1995} presented an anisotropic damage model 
where damage initiation is controlled by the failure criteria developed by \cite{Hashin1980}. \cite{Chatiri2013} extend this model to the three-dimensional case and 
incorporate damage initiation based on the Puck failure criterion. A model with initiation criteria based on the failure mode concept of Cuntze is presented in \cite{Boehm2010a} 
for woven or braided reinforcements and in \cite{Boehm2011} for non-crimp fabrics.

Element erosion can be triggered by a critical value of the damage variable. These damage models are coupled to a generally anisotropic material model that represents the deformation behaviour 
in the undamaged state. Depending on the composition of the FRP material and the geometrical arrangement of the reinforcing fibres, purely elastic, viscoelastic or viscoplastic models can be 
used, see e.g. \cite{Kaestner2012} for a classification of characteristic phenomena of the mechanical material behaviour.

This paper is outlined as follows: In Section \ref{sec:Material}, the constitutive model used to represent the mechanical behaviour of the composite is outlined.
% In particular, an anisotropic viscoelastic model is presented which accounts for strain rate dependent effects, creep, and relaxation which can be observed in FRP for matrix dominated loading directions.
In particular, a continuum damage model that allows the description of intra-laminar damage in FRP materials and its implementation in a commerical finite element code are presented. Section \ref{sec:Identification} addresses the identification of the associated material parameters. In addition to experiments for the identification of intra- and interlaminar damage, we demonstrate the potential of a multi-scale approach that predicts the effective behaviour of the laminates based on the properties of the used polymeric matrix and the fibre material. The developed modelling approach is eventually applied to simulate an SPR process to join an FRP laminate with an aluminium sheet in Section \ref{sec:Simulation}. The paper is closed by concluding remarks and an outlook to necessary modifications in Section \ref{sec:Conclusion}.

\section{Constitutive modelling of fibre reinforced polymers}
\label{sec:Material}
According to the authors' experience, the effective mechanical behaviour of FRP materials is governed by characteristic damage processes and viscoelastic effects which include relaxation and creep phenomena that can impair the long-term performance of structural components.  
In this contribution we apply a homogenisation scheme that allows the computation of effective viscoelastic properties of composites and provides the elastic stiffness as a limiting case. While the quasistatic viscoelastic behaviour might be of interest for the analysis of the joint strength especially for thermoplastic matrices, we assume that damage effects dominate the characteristics of the SPR processes. An initially elastic damage model is therefore employed within the process simulation in Section \ref{sec:Simulation}.
It has been implemented into the explicit version of ABAQUS in terms of a user-defined material. All multi-axial constitutive relations are given in vector-matrix notation as typically used in FE codes.
 
\subsection{Continuum damage}
\label{sec:Identification_damage}
To capture the characteristic damage phenomena within the process simulation, an orthotropic continuum damage model is utilised. The definition of the material model comprises three fundamental ingredients:
\begin{enumerate}
  \item A stress-strain relation that accounts for damage in the material,
  \item Conditions if an evolution of damage may occur, and
  \item Evolution equations that define the temporal change of the damage variables.
\end{enumerate}
The formulation of the used model is based on the effective stress concept by \cite{Kachanov1958}, the evolution equations introduced by \cite{Matzenmiller1995} for the two-dimensional case, and their generalisation to the three-dimensional case by \cite{Chatiri2013}. For brevity we only present the used constitutive relations without details of their derivation. A more comprehensive presentation of the damage model can be found in \cite{Chatiri2013}.

As a consequence of damage, the stiffness of the material is reduced. A stress-strain relation that accounts for the stiffness degradation of the damaged material is given by
\begin{equation}
\label{eq:stress_strain_damage}
 \vect{\sigma} = \hat{\matr{C}}\vect{\varepsilon}= \left( \hat{\matr{H}} \right)^{-1}\vect{\varepsilon}~\text{.}
\end{equation}
It relates the nominal stress $\vect{\sigma}$ to the strain $\vect{\varepsilon}$ in terms of the degraded stiffness matrix $\hat{\matr{C}}$ or the compliance matrix $\hat{\matr{H}}$, respectively.  It can be observed from the structure of the compliance matrix
\begin{equation}
 \textstyle\hat{\matr{H}}=\begin{bmatrix}
    \frac{1}{(1-\omega_{11})E_{11}}        & -\frac{\nu_{21}}{E_{22}} & -\frac{\nu_{21}}{E_{22}} & 0     & 0     & 0   \\[0.6ex]
   -\frac{\nu_{12}}{E_{11}} &\frac{1}{(1-\omega_{22})E_{22}}        & -\frac{\nu_{32}}{E_{22}} & 0     & 0     & 0   \\[0.6ex]
   -\frac{\nu_{12}}{E_{11}} & -\frac{\nu_{23}}{E_{22}} &  \frac{1}{(1-\omega_{33})E_{22}}        & 0    & 0     & 0   \\[0.6ex]
   0                    & 0                    & 0                    & \frac{1}{(1-\omega_{23})G_{23}}  & 0 & 0 \\[0.6ex]
   0                    & 0                    & 0                    & 0     & \frac{1}{(1-\omega_{31})G_{12}}  & 0 \\[0.6ex]
   0                    & 0                    & 0                    & 0     & 0     &  \frac{1}{(1-\omega_{12})G_{12}}
\end{bmatrix}
\end{equation}
that damage is accounted by six scalar damage variables $\omega_{ij}\in [0,1]$. 

The evolution of the damage variables from the undamaged initial state $\omega_{ij}=0$ to complete damage indicated by $\omega_{ij}=1$ is governed by the definition of damage surfaces. Two principal damage modes, i.e. fibre and matrix failure as illustrated in \reff{fig:Methodical_Framework}, are considered according to the failure criterion introduced by \cite{Puck1998}. All damage variables are related to these two modes. The failure surface
\begin{equation}
f_{\parallel}^{\pm}=\left( \frac{\hat\sigma_{11}}{R^\pm_\parallel} \right)^2 -r_\parallel^\pm=0 
\label{eq:schaedigungsflaeche-spannung-effektiv-1}
\end{equation}
controls damage in the fibre $(\cdot)_\parallel$ mode with the effective stress in the fibre direction $\hat\sigma_{11}$, fibre strength $R_\parallel^\pm$ and a threshold $r_\parallel^\pm \geq 1$ for damage initiation and propagation. Matrix failure $(\cdot)_\perp$ 
for stresses $\sigma_n\geq 0 $ is governed by
\begin{equation}
\begin{split}
  \textstyle f_\perp^+\textstyle =\sqrt{\left[ \frac{1}{R^+_\perp}- \frac{p^+_{\perp \Psi}}{R^A_{\perp \Psi}} \right]^2 \sigma^2_n + \left[ \frac{\tau_{nt}}{R^A_{\perp\perp}} \right]^2+ \left[\frac{\tau_{n1}}{R_{\perp\parallel}} \right]^2}+\frac{p^+_{\perp \Psi}}{R^A_{\perp \Psi}}\sigma_n-r^+_\perp=0
  \label{eq:schaedigungsflaeche-spannung-effektiv-2}
\end{split}
\end{equation}
and for $\sigma_n < 0$ by
\begin{equation}
  \textstyle f_\perp^-\textstyle =\sqrt{\left[ \frac{p^-_{\perp \Psi}}{R^A_{\perp \Psi}} \sigma_n \right]^2 + \left[ \frac{\tau_{nt}}{R^A_{\perp\perp}} \right]^2+ \left[\frac{\tau_{n1}}{R_{\perp\parallel}} \right]^2}+\frac{p^-_{\perp \Psi}}{R^A_{\perp \Psi}}\sigma_n-r^-_\perp=0
  \label{eq:schaedigungsflaeche-spannung-effektiv-3}
\end{equation}
with $R^\pm_{\perp}$ and $R_{\perp\parallel}$ the strength values for normal and shear load cases, respectively, and the threshold for matrix damage $r^\pm_{\perp}\geq 1$.  The superscript $(\cdot)^\pm$ indicates quantities which can have different values 
for tension and compression loadings. The stresses $\sigma_n(\theta_\text{fp})$, $\tau_{nt}(\theta_\text{fp})$ and $\tau_{n1}(\theta_\text{fp})$ in Equations \refe{eq:schaedigungsflaeche-spannung-effektiv-2} and \refe{eq:schaedigungsflaeche-spannung-effektiv-3} represent 
the projected stress components of the effective stress $\hat{\vect\sigma}$ on a fracture plane that is orthogonal to the fibre direction. The orientation of the fracture plane is fully determined by the rotation angle $\theta_\text{fp}$ which is 
calculated from the condition $\left[f^\pm_\perp(\theta)\right]_\text{max}=f^\pm_\perp(\theta_\text{fp})$, i.e. the fracture plane which generates the greatest failure effort is chosen. Further details of this so called action plane concept and the corresponding 
transformations can be found in \cite{Puck2002}.

The ratios $\frac{p^\pm_{\perp\Psi}}{R^A_{\perp\Psi}}$ that occur in Equations \refe{eq:schaedigungsflaeche-spannung-effektiv-2} and \refe{eq:schaedigungsflaeche-spannung-effektiv-3} are defined by 
\begin{equation}
 \frac{p^\pm_{\perp \Psi}}{R^A_{\perp \Psi}}=\frac{p^\pm_{\perp\perp}}{R^A_{\perp\perp}} \frac{\tau^2_{nt}}{\tau^2_{nt}+\tau^2_{n1}}+\frac{p^\pm_{\perp\parallel}}{R_{\perp\parallel}} \frac{\tau^2_{n1}}{\tau^2_{nt}+\tau^2_{n1}}
\end{equation}
with the parameter 
\begin{equation}
 R^A_{\perp\perp}=\frac{R^-_\perp}{2(1+p^-_{\perp\perp})}
\end{equation}
and the Puck constants $p^\pm_{\perp\perp}$, $p^\pm_{\perp\parallel}$. 

The current state of the individual damage variables is given by
\begin{equation}
 \vect{\omega} = \sum \limits_{I=\parallel,\perp}  \phi_I \, \vect{q}_I
\end{equation}
with the scalar valued growth functions for each damage mode \citep{Chatiri2013}
\begin{align}
\phi_I=%\int\limits_0^t \dot\phi_I \,\text{d}\overline t=
1-e^{\frac{1}{m_I}(1-r_I^{m_I})}~\text{.}
\end{align}
The growth functions involve the softening parameter $m_{I}$, which controls the shape of the evolution function, as well as the damage threshold $r_I\geq 1$. The parameters will be identified from experimental results in Section \ref{sec:Identification}.

The coupling vectors $\vect{q}_I$ link the individual damage variables which are the entries of $\vect{\omega}$ to the fibre and matrix damage mode, respectively. It is assumed in the remainder of this work, that the fibre mode only degrades the stiffness in the fibre direction modelled by $\omega_{11}$, while damage in the matrix mode leads to increasing values of $\omega_{i\neq 1, j\neq 1}$, i.e.
\begin{align}
  \vect{q}_{\parallel}&=\left[\begin{array}{cccccc} 1 & 0 & 0 & 0 & 0 & 0\end{array}\right]^T\\
  \vect{q}_{\perp}&=\left[\begin{array}{cccccc} 0 & 1 & 1 & 1 & 1 & 1\end{array}\right]^T~\text{.}
\end{align}

\begin{figure}
  \begin{center}
    \includegraphics[scale=1]{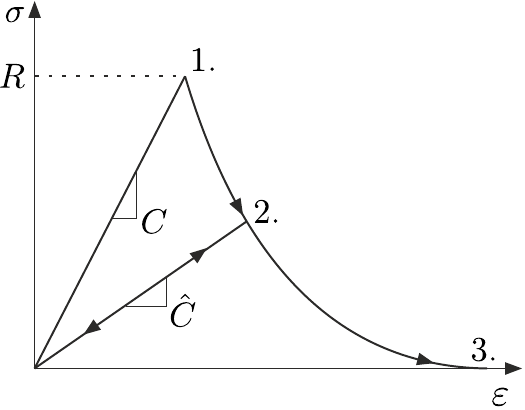}
    \caption{Uniaxial stress-strain curve and effective stiffness simulated with the outlined continuum damage model.}
  \label{fig:Stiffness_degradation}
  \end{center}
\end{figure}

A typical uniaxial stress-strain curve simulated with the presented damage model is illustrated in \reff{fig:Stiffness_degradation}. Starting from the stress and strain free initial state, the material follows a linear stress-strain relation with the initial stiffness ${C}$. The condition for the initiation of damage specified by the damage surface is met at point 1. This results in a growing damage variable and a degradation of the stiffness $\hat{C}$. During the un- and reloading from point 2 on, the damage surface prevents any changes of the damage variable and hence the stiffness remains constant. A subsequent increase of the loading results in the complete damage of the material ($\omega=1$) at point 3.

\subsection{Implementation}

The damage model described in Section \ref{sec:Identification_damage} has been implemented in ABAQUS explicit. The user subroutine VUMAT enables the definition of mechanical material models which are not readily available in the programme \citep{abaqus}. This subroutine is called for block of integration points and has to provide updates of the stress tensor and state variables during a nonlinear finite element analysis.

The experimental results of Section \ref{sec:Identification} indicate that failure of the composite occurs at small strains. However, the self-piercing rivetting process will induce large deflections of the laminae which lead to a rotation of the principal axes of the material relative to the fixed global coordinate system. From a finite element point of view, the simulation of the behavior of the composite is a small strain - large rotation problem which is handeled efficiently by a corotational formulation \citep{Belytschko2000}. This procedure is activated by the NLGEOM option. In ABAQUS the rotations are e.g. obtained from the polar decomposition of the deformation gradient $\mathbf{F}=\mathbf{R}\cdot\mathbf{U}$ where $\mathbf{R}$ is an orthogonal tensor representing rigid body rotations and $\mathbf{U}$ is the stretch tensor that describes the deformation. The key feature of corotational formulations is a Carthesian coordinate system $\hat{\mathbf{e}}_i,~i=1,2,3$ which rotates with the body, i.e. $\hat{\mathbf{e}}_i=\mathbf{R}\cdot\mathbf e_i$.

The user subroutine VUMAT provides strains, stretches, and state variables with respect to the rotated frame.
 With the strains in the corotational frame $\hat{\varepsilon}_{ij}$ being small, the constitutive model of Section \ref{sec:Identification_damage} can be employed directly to compute the updated corotational stress coordinates $\hat{\sigma}_{ij}$ according to the stress-strain relation \refe{eq:stress_strain_damage}. It is noted, that the time rate of the corotational stress coordinates is equivalent to the \textsc{Green-Nagdi} rate of the \textsc{Cauchy} stress
 \begin{equation}
 \label{eq:stress_rate}
 \overset{\nabla}{\boldsymbol \sigma}=\dot{\boldsymbol \sigma}
 -\dot{\mathbf{R}}\cdot\mathbf{R}^{\text T} \cdot \boldsymbol\sigma
 +\boldsymbol\sigma \cdot \dot{\mathbf{R}}\cdot\mathbf{R}^{\text T}~.
 \end{equation}
 Updates of the \textsc{Cauchy} stress coordinates with respect to the fixed global coordinate system, which are required to update the internal force vector, can be obtained from Equation \refe{eq:stress_rate}.

\begin{figure}
  \begin{center}
    \includegraphics[width=1\textwidth]{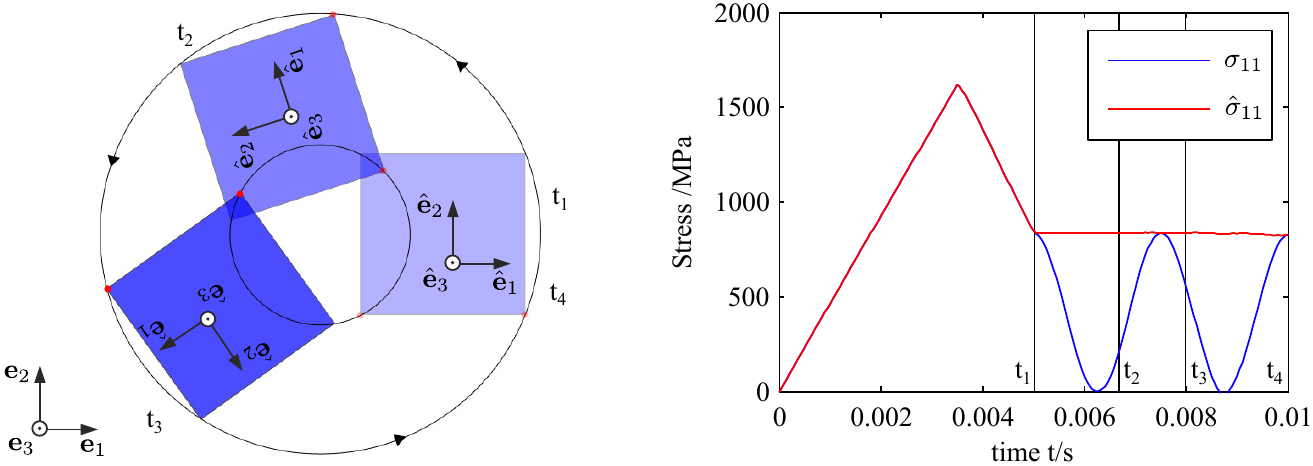}
    \caption{Corotational finite element formulation for the small strain - large rotation problem: comparison of global ($\sigma_{11}$) and corotational stress ($\hat\sigma_{11}$) coordinates computed from the defined material model during finite rotation of a single element.}
  \label{fig:rot_element}
  \end{center}
\end{figure}

Figure \ref{fig:rot_element} provides an illustrative example to verify the implementation. In this problem, a displacement controlled loading is applied to a single element for $0\le t\le t_1$. It can be seen from the uniaxial stress-strain curve in \reff{fig:rot_element} that the stress increases linearly until degradation is initiated. For times $t>t_1$ the mechanical load is kept constant and a rigid body rotation is prescribed to the element. As a consequence of the rotation, the stress with respect to the fixed global coordinate system starts to oscillate, i.e. the material time derivative of the $\textsc{Cauchy}$ stress is not objective. However, the corotational stress coordinate $\hat\sigma_{11}$ with respect to the rotating local frame is constant as expected, because the mechanical load of the element does not change. Equivalently, $\overset{\nabla}{\sigma}_{11}=0 $ holds for the objective \textsc{Green-Nagdi} rate of the \textsc{Cauchy} stress.
\vspace{12pt}
\vspace{36pt}
\section{Parameter identification}
\label{sec:Identification}

In this section, the required parameters of the anisotropic damage model and the cohesive zone model are to be identified. It has to be noted, that the constitutive model introduced above involves a considerable number of material parameters which is a result of the complexity of the heterogeneous material, and the damage phenomena that have to be modelled. The identification of all parameters will consequently lead to an extensive experimental effort. Moreover, the broad range of reinforcing fibres, matrix materials, and textile reinforcements require these experiments to be repeated for any change in the composition of the material.

Therefore, the experimental characterisation of the FRP material is partly replaced by a numerical  homogenisation procedure. That is, the undamaged effective material behaviour is computed from the properties of matrix and fibres.  
As the prediction of the effect of microscopic damage on the overall mechanical response is a topic of ongoing research, e.g. the influence of interface failure on the macroscopic stress-strain repsonse has been analysed in \cite{Kaestner2015}, the parameterisations of the damage model, Section \ref{sec:Damage_Composite}, and the cohesive zone model in Section \ref{sec:Cohesive} are still based on experiments.

\subsection{Prediction and validation of the effective response of FRP}
\label{sec:Homogenisation_Composite}
In the following, the effective material behaviour of a lamina will be determined numerically using a homogenisation technique. 
The major advantage of this approach is that once the behaviour of a set of constituents has been investigated, it is possible to predict the effect of different material combinations and various types of fibre reinforcements for load cases that are hard to realise in experiments. 

\begin{figure}
  \begin{center}
    (a)~\includegraphics[scale=1.25]{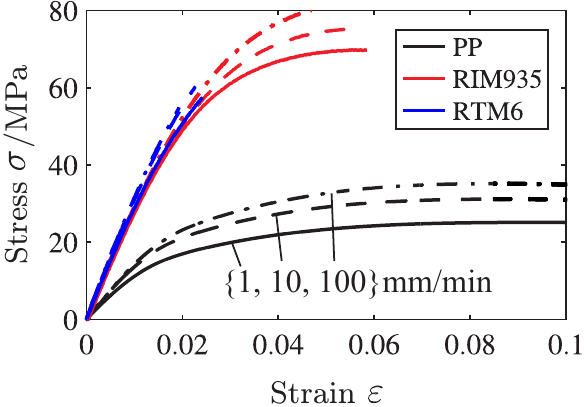}\qquad
    (b)~\includegraphics[scale=1.25]{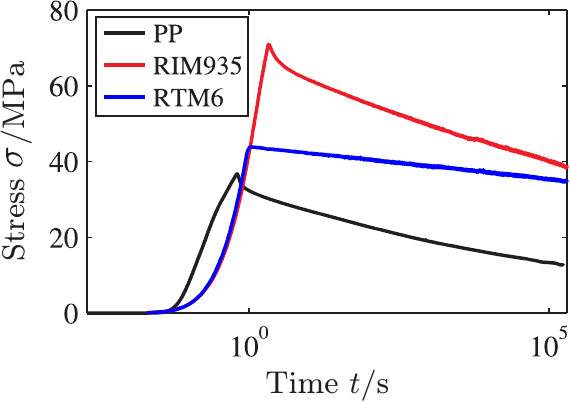}
    \caption{Monotonic uniaxial tests carried out for the epoxy resins RIM935, and RTM6, and the thermoplastic Polypropylene \citep{Kaestner2010a}: (a) stress-strain curves obtained in tensile tests at three distinct strain rates, and (b) stress-time curves of relaxation experiments.}
    \label{fig:TestsPolymer}
  \end{center}
\end{figure}

While common fibre materials are either isotropic or transversely isotropic elastic materials, typical polymeric matrices exhibit a non-linear, rate dependent deformation behaviour as shown in \cite{Kaestner2012}. Exemplary stress-strain curves for three widely used polymeric matrices, the epoxy resins RIM935 and RTM6, and the thermoplastic Polypropylene (PP), are illustrated in \reff{fig:TestsPolymer}~(a) at three distinct displacement velocities $v=\{\num{1},\num{10}, \num{100}\}\SI{}{\milli\meter\per\minute}$. We therefore apply a viscoelastic homogenization procedure. This technique produces effective elastic stiffness properties as limiting case.

Whether viscoelastic effects have to be taken into account in the simulation of an SPR process and the subsequent numerical analysis of the joint strength hinges on the applied materials. If unreinforced polymers are used in SPR processes as in e.g. \cite{Zhang2014}, the high strain rates during the SPR process will result in a significant change of the material behaviour compared to quasi-static experiments. In addition, considerable stress relaxation and creep phenomena are to be expected after the setting process. 
For FRP the short-term behaviour is clearly dominated by the fibres, but stress relaxation could be an issue especially for fibre reinforced thermoplastics, cf. \cite{Kaestner2011}.

\subsubsection{Parameter identification for RIM935}
\label{ssec:Polymer_param}

A generalised Maxwell model \citep{Tschoegl1989} is employed to describe the viscoelastic behaviour. For the case of linear viscoelasticity considered here, an explicit representation of the stress response of the model for the initial conditions $\sigma(t=0)=0$ and $\varepsilon(t=0)=0$ is given by the convolution integral, also known as the Boltzmann superposition integral
\begin{equation}
 \sigma (t) = \int\limits_0^t G(t-\xi)\dot{\varepsilon}(\xi)~d\xi \quad \text{with} \quad G(t)=E+\sum_{I=1}^Nc_I e^{-\frac{t}{\tau_I}}
 \label{eqn:Sol_Explicit}
\end{equation}
where $\dot{\varepsilon}(\xi)=\frac{d \varepsilon}{d \xi}$ is the strain rate history of the deformation process and $G(t)$ is the relaxation function of the generalised Maxwell model. The elastic modulus $E$ and the sets of relaxation times $\vect{\tau}=\{\tau_I\}_{I=1}^N$ and relaxation strengths $\vect{c}=\{c_I\}_{I=1}^N$ of each $N$ Maxwell elements, fully describe the uniaxial viscoelastic material behaviour in terms of the relaxation function $G(t)$.

\begin{figure}
  \begin{center}
    \includegraphics[scale=1]{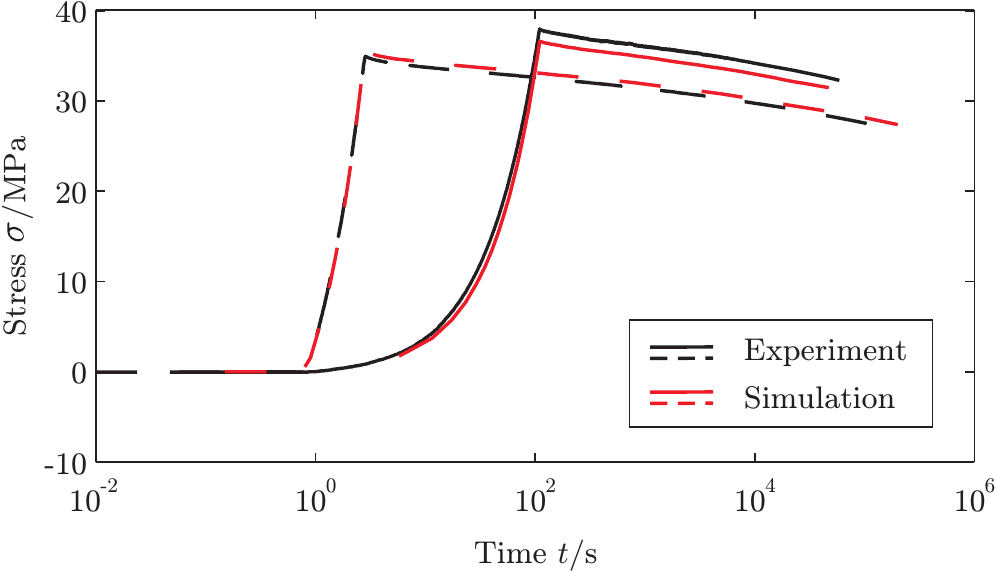}
    \caption{Simulation of uniaxial relaxation tensile tests at $\bar{\varepsilon}=\SI{1,5}{\percent}$ and comparison with experimental results.}
  \label{fig:Simulation_1D_relaxation}
  \end{center}
\end{figure}

The identification of the parameters for RIM935, which is used as matrix material in this study, is accomplished by a systematic, recursive analysis of the stress-time curve obtained in relaxation experiments \citep{Emri1993,Kaestner2012}. For an a priori given number of $N=8$ Maxwell elements with relaxation times
\begin{equation}
  \vect{\tau}= \left[ 10^{-1} ~~10^0 ~~10^1 ~~10^2 ~~10^3 ~~10^4 ~~10^5 ~~10^7 \right]^T \SI{}{\second}
  \label{eq:relaxation_times}
\end{equation}
the following set of relaxation strengths
\begin{equation}
 \vect{c}= \left[ 90.4 ~~94.5~~66.0~~71.1~~83.3~~12.9~~13.8~~1820 \right]^T \SI{}{\mega\pascal} 
 \label{eq:relaxation_strengths}
\end{equation}
has been identified from the stress-time curves of two relaxation tests performed at $\bar{\varepsilon}=\SI{1,5}{\percent}$. \reff{fig:Simulation_1D_relaxation} shows the comparison of the results from  relaxation experiments and the model response. It can be seen that the model is capable of describing the stress relaxation for both, the loading path and the subsequent stress relaxation, with good accuracy.
Presuming isotropic material behaviour of the polymer, the multi-axial generalisation of Equation \refe{eqn:Sol_Explicit} to 
\begin{equation}
 \vect{\sigma} (t) = \int_{0}^t \matr{G}(t-\xi) \dot{\vect{\varepsilon}}(\xi)~d\xi\quad \text{with} \quad
 \matr{G}(t)=\matr{G}^{\text{eq}}+\sum_{I=1}^N\matr{G}_I e^{-\frac{t}{\tau_I}}
 \label{eq:relaxation_function}
\end{equation} 
is achieved by the definition of the Poisson ratio. Here, a constant value of $\nu=\num{0,4}$ is assumed.
\subsubsection{Homogenisation}
\label{ssec:Homogenisation_Composite}
\begin{figure}
  \begin{center}
    (a)~\includegraphics[scale=1.0]{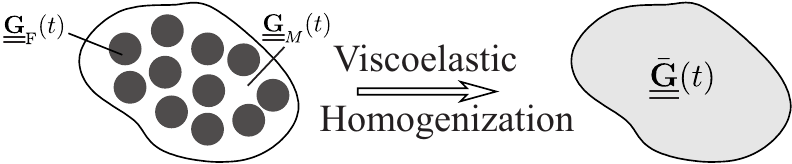}\qquad
    (b)~\includegraphics[scale=1.0]{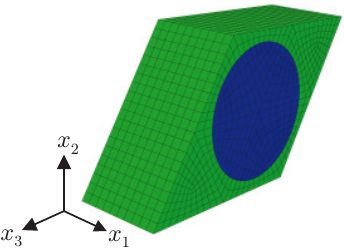}
    \caption{Homogenisation for viscoelastic materials: (a) The effective relaxation function $\bar{\matr{G}}(t)$ is computed based on the corresponding functions of the matrix $\matr{G}_{\,\text{M}}(t)$ and the fibres $\matr{G}_{\,\text{F}}(t)$. The geometrical arrangement of the fibres has to be represented in terms of an RVE. For unidirectional reinforcements a rhombic unit cell (b) results in the expected transversely isotropic material behaviour.}
  \label{fig:Homogenisation_scheme}
  \end{center}
\end{figure}

The standard homogenisation procedures developed for linear elasticity \cite{Hill1963} can be generalised to the case of linear viscoelasticity based on an 
elastic-viscoelastic correspondence principle \citep{Pierard2004}. Assume the material behaviour of the matrix $\matr{G}_{\,\text{M}}(t)$ and the fibres $\matr{G}_{\,\text{F}}(t)$ to be given in terms of the relaxation function \refe{eq:relaxation_function}, \reff{fig:Homogenisation_scheme}~(a).  
The rhombic RVE in \reff{fig:Homogenisation_scheme} (b) is equivalent to the hexahedral unit cell used e.g. in \cite{Kaestner2011} and will result in a transversely isotropic effective material behaviour. 

The viscoelastic homogenisation procedure then allows to predict the effective properties of an orthotropic viscoelastic relation given by the convolution integral
\begin{equation}
  \bar{\vect{\sigma}} (t) = \int \limits_{0}^t \bar{\matr{G}}(t-\xi) \dot{\bar{\vect{\varepsilon}}} ~d\xi
  \label{eq:viscoel_ortho_stress_strain}
\end{equation}
and the effective orthotropic relaxation function
\begin{equation}
 \bar{\matr{G}} (t) = \bar{\matr{G}}_\infty + \sum \limits_{I=1}^{\bar{N}} \bar{\matr{G}}_I e^{-\frac{t}{\tau_I}}
 \label{eq:relaxation_function_eff}
\end{equation}
where $\bar{N}$ is the number of Maxwell elements required to represent the macroscopic behaviour. 

\begin{figure}
  \begin{center}
    \includegraphics[scale=1]{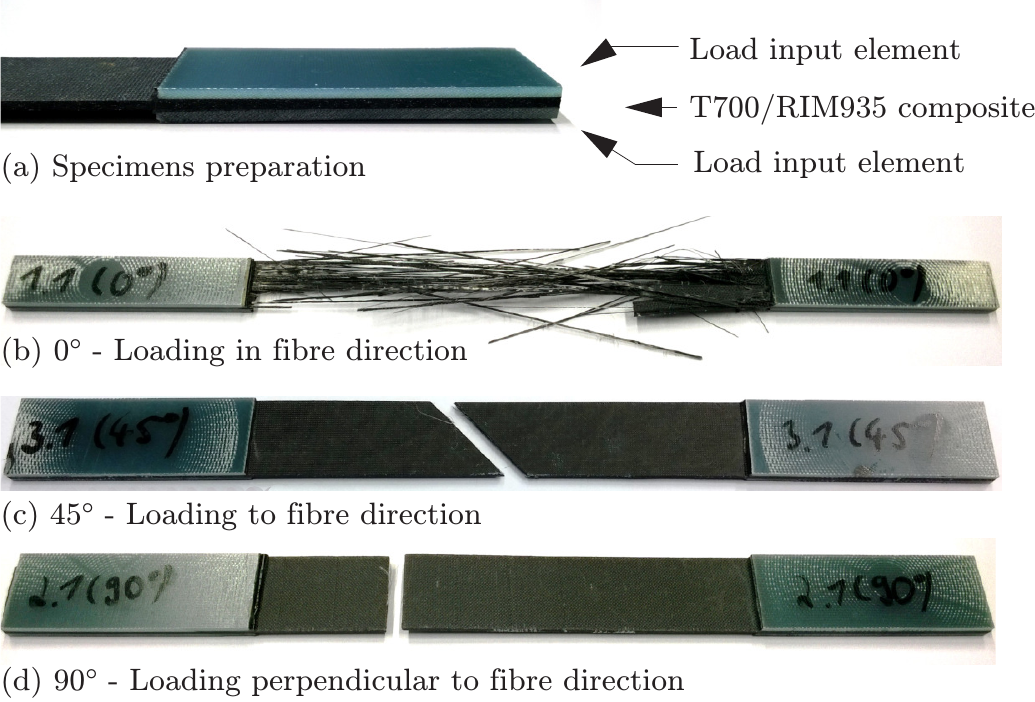}
    \caption{Unidirectionally reinforced test specimens for tensile tests: (a) composite specimen load input elements, (b) \ang{0}-loading in fibre direction, (c) \ang{45}-loading, and (d) \ang{90}-loading to fibre direction - specimens after quasi static tension.}
  \label{fig:Exp_tension_composite}
  \end{center}
\end{figure}

An experimental validation of the homogenisation procedure is obtained from the comparison of the effective response computed for the T700/RIM935 composite with experimental results for loading in the fibre direction, and loading perpendicular to the fibre direction according DIN ISO 527-5, cf. \reff{fig:Exp_tension_composite}. It can be seen that the predicted effective stress-strain curves are in good agreement with the mechanical response observed in the experiment, \reff{fig:Sim_Exp_CF_RIM935_Damage}. Especially the large difference between the two loading directions is captured accurately. 

As there are no significant non-linearities in both stress-strain curves, the effective elastic material behaviour can be represented in terms of an effective elastic stiffness matrix, e.g. the instantaneous stiffness $\matr{C}=\matr{G}^{\text{eq}}+\sum_{I=1}^{\bar{N}=8}\matr{G}_I$ can be used for the anisotropic continuum damage model. From the viscoelastic homogenisation procedure based on the properties of RIM935 and the carbon fibre T700, the stiffness 
\begin{equation}
\matr{C}=\begin{bmatrix}
          119376    &    5033      &      5033              &           0      &                   0         &                0\\
           5033      &    10273     &    6040                &         0        &                 0           &              0\\
           5033      &      6040      &  10273               &          0        &                 0           &              0\\
                         0               &          0                &         0        &    2116              &           0           &              0\\
                         0               &          0                &         0           &              0       &    2561         &               0\\
                         0                &         0                &         0            &             0          &               0     &      2562
\end{bmatrix}\text{MPa}
\end{equation}  
is obtained for a fibre volume fraction of $v_{\text{F}}=\num{0,5}$. For other material combinations, e.g. glass fibres and thermoplastic matrices as investigated in \cite{Kaestner2011}, more pronounced non-linearities are observed. 

\subsection{Identification of parameters for the damage model}
\label{sec:Damage_Composite}

In the following, the parameters of the damage model are to be identified from experiments on a T700/RIM935 composite with unidirectional reinforcement. The set of material parameters includes strength values for different loadings $R_{\parallel}^\pm, R_{\perp\perp}^\pm, R_{\perp\parallel}$, the Puck parameters $p^\pm_{\perp\parallel}, p^\pm_{\perp\perp}$ and the parameters $m_I$ which control the damage propagation. In order to identify all parameters, a comprehensive programme of in-plane and out-of-plane tension and compression as well as shear experiments would have to be carried out. However, not all of these experiments can be realised or will produce reasonable results. In this study, only in-plane tensile tests, $0^\circ, 45^\circ$ and $90^\circ$ to the fibre direction, as well as out-of-plane compression tests have been available. Consequently, several assumptions and simplifications have to be made.

\subsubsection{In-plane tensile tests}
\begin{figure}
  \begin{center}
    (a)~\includegraphics[scale=1]{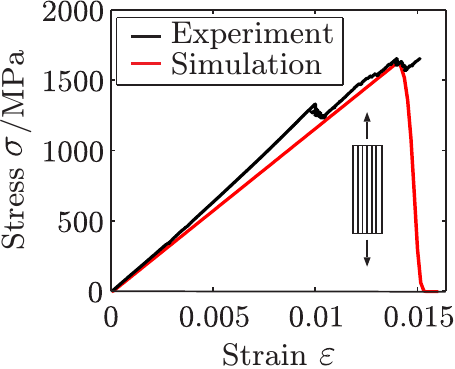}
    (b)~\includegraphics[scale=1]{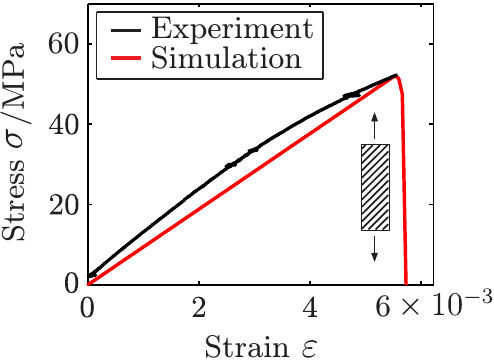}
    (c)~\includegraphics[scale=1]{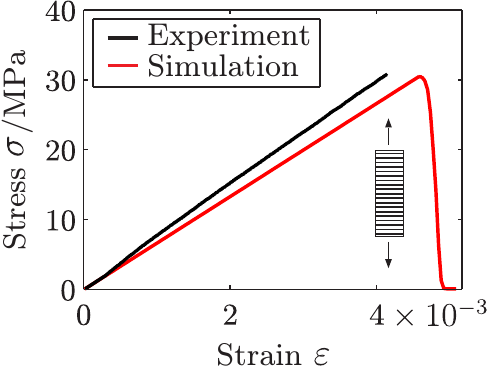}
    \caption{Comparison of the mechanical response simulated with the orthotropic continuum damage model to experimental results: (a) \ang{0}-loading in fibre direction, (b) \ang{45}-loading, and (c) \ang{90}-loading to the fibre direction.}
  \label{fig:Sim_Exp_CF_RIM935_Damage}
  \end{center}
\end{figure}

The in-plane tensile tests reveal a brittle behaviour in both principal loading directions, \reff{fig:Sim_Exp_CF_RIM935_Damage}, which is governed by fibre rupture and interface failure, respectively. These characteristics can be recovered with the damage model for exponents $m_I\gg 1$. Here, values of $m_{\parallel}^+=\num{30}$, and $m_{\perp}^+=\num{90}$ have been chosen. Due to the uniaxial stress states in the specimens, the tensile strengths $R_{\parallel}^+=\SI{1627}{\mega\pascal}$, and $R_{\perp}^+=\SI{32}{\mega\pascal}$ can be read directly from the corresponding stress-strain curves. The strength $R_{\perp\parallel}$ was varied during several simulations for uniaxial tension in $45^\circ$ to fibre direction. $R_{\perp\parallel}=\SI{27}{\mega\pascal}$ has been adapted to achieve a good agreement between the experimental and the numerical results. Figure \ref{fig:Sim_Exp_CF_RIM935_Damage} shows the comparison of the experimentally and numerically obtained stress-strain curves for all three loadings and demonstrates a good agreement between simulation and experiment.

\subsubsection{Out-of-plane compression tests}
\begin{figure}
  \begin{center}
    (a)~\includegraphics[scale=2.0]{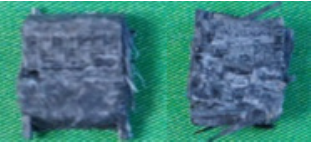}\qquad
    (b)~\includegraphics[]{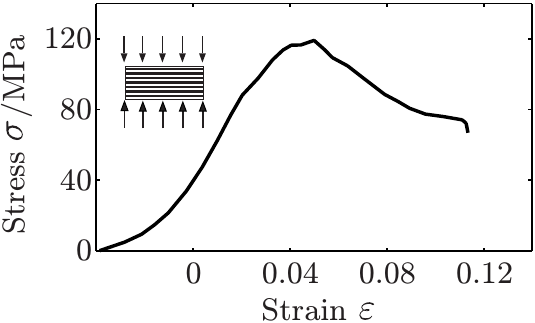}
    \caption{Out-of-plane compression loading of a $\SI{0}{\degree}$ composite: (a) tested specimens, and (b) experimental results.}
  \label{fig:CompTest}
  \end{center}
\end{figure}

Before cutting the fibres of the composite, the rivet exerts compressive loads on the material. To determine the corresponding parameters of the damage model, out-of-plane compression tests have been carried out. Compressive stresses $\sigma_{33}<0$ will primarily lead to matrix crushing. A unidirectional composite consequently loses its structural integrity which results in low strength parameters. Experimental data for a $\SI{0}{\degree}$ lay-up have been analysed to identify the parameters for out-of-plane loading. Nevertheless, the composite still shows a more ductile force-displacement behaviour compared to tensile loadings, \reff{fig:CompTest}~(b). The parameters $R_{\perp}^-=\SI{123}{\mega\pascal}$ and $m_{\perp}^-=\num{7}$ account for a proper representation of the compression behaviour.
\subsubsection{Assumptions}
The available experiments do not allow for a complete and consistent identification of all material parameters. As pure in- and out-of-plane shear as well as in-plane compression loadings are hard to realise experimentally, the following assumptions are made:
\begin{enumerate}
  \item The Puck constants $p^-_{\perp\parallel}=0.25, p^+_{\perp\parallel}=0.3$ and $p^\pm_{\perp\perp}=0.25$ were chosen in according to \cite{Puck2002b}.
  \item The parameters for compression loading in the fibre direction, i.e. $R_{\parallel}^-$ and $m^-_{\parallel}$ are estimated because of missing experimental data. Using the tensile parameters for compression mode would strongly overestimate the compression behaviour because failure under compressive loads is driven by buckling rather than fibre breakage. With $R_{\parallel}^-=\SI{1000}{\mega\pascal}$ the compression strength equals $60 \%$ of the tensile strength following \cite{Puck2002}.
\end{enumerate}
The complete set of used material parameters is summarised in \reft{tab:MPDamage}. 
\begin{table}
\centering
\caption{Material parameters of the damage model.}
\begin{tabular}{p{2.5cm}ccccccc}
   \hline\hline
   Load Case & $R_{\parallel}$/MPa & $R_{\perp}$/MPa & $R_{\perp\parallel}$/MPa & $m_{\parallel}$ & $m_{\perp}$ & $p_{\perp\perp}$ & $p_{\perp\parallel}$ \\\hline
   Tension~(+) & 1627 & 32 & 27 & 30 & 90 & 0.25 & 0.3 \\
   Compress.~(-) & 1000 & 123 & 27 & 30 & 7 & 0.25 & 0.25 \\\hline\hline
\end{tabular}\\[6pt]
\label{tab:MPDamage}
\end{table}

\subsection{Cohesive Zone}
\label{sec:Cohesive}
The delamination of the individual composite layers is captured by the insertion of a cohesive zone model in between the individual laminates. Therefore the surface-based cohesive model of ABAQUS is utilised. 
The mechanical response is defined in terms of a bilinear traction-separation law, cf. \reff{fig:Methodical_Framework}
\begin{equation}
\label{eq:tsl}
\vect{t}_\text{c}=
\begin{bmatrix}
t_\text n\\
t_\text s\\
t_\text t
\end{bmatrix}
=(1-D_\text{CZ})
\begin{bmatrix}
k_\text n & 0 & 0 \\
0 & k_\text s & 0 \\
0 & 0 & k_\text t
\end{bmatrix}
\begin{bmatrix}
\langle\delta_\text n\rangle_+\\
\delta_\text s\\
\delta_\text t
\end{bmatrix}
+k_\text n
\begin{bmatrix}
\langle\delta_\text n\rangle_-\\
0\\
0
\end{bmatrix},
\end{equation}
with $t_I$ the tractions, $\delta_I$ the separations and $k_I$ the elastic stiffness parameters. The indices $I=\text n,\text t,\text s$ correspond to the normal and the two tangential directions of the interface, respectively. 
It is noted that the normal stiffness $k_\text n$ is only degraded for positive separations. For negative normal separations it remains unchanged, 
i.e. it acts as a penalty stiffness to prevent penetration of the composite layers. In Equation \refe{eq:tsl} this case is fulfilled through the Macaulay brackets defined by $\langle x\rangle_\pm=\frac{1}{2}(x\pm|x|)$. 

The initiation of interfacial damage is controlled by the stress criterion 
\begin{equation}
\label{es:maxs}
 \text{max}\left( \frac{\langle t_\text n\rangle_+}{t_\text n^0},\frac{t_\text s}{t_\text s^0},\frac{t_\text t}{t_\text t^0} \right)-1=0
\end{equation}
which compares the actual stress state with the cohesive strengths $t^0_I$. 

The description of the damage evolution is based on the definition of the effective scalar separation \linebreak$\delta=\sqrt{\langle \delta_\text n \rangle_+^2 +\delta_\text s^2+\delta_\text t^2}$ and a linear degradation of the form
\begin{equation}
 D_\text{CZ}=\frac{2G_\text{c}}{\delta_\text{max}}\frac{\delta_\text{max}-\delta_0}{2G_\text{c}-t_0\delta_0},
\end{equation}
with $\delta_\text{max}=\max\limits_{\tau\leq t}[\delta(\tau)]$ the maximum of the effective separation and $t_0$ and $\delta_0$ the effective traction and separation at damage initiation. 
The critical mixed mode energy release $G_\text{c}$ is based on a simple power law criterion and is computed from
\begin{equation}
 G_\text{c}=\frac{G_\text{n}+G_\text{s}+G_\text{t}}{\frac{G_\text{n}}{G_\text{Ic}}+\frac{G_\text{s}}{G_\text{sc}}+\frac{G_\text{t}}{G_\text{tc}}}.
\end{equation}
The critical energy release rates $G_{i\text c}$ has to be determined experimentally and the actual energy release rates $G_i$ represent the work done by the tractions and separations in the individual modes. 

\begin{figure}
  \begin{center}
    (a)~\includegraphics[scale=1.0]{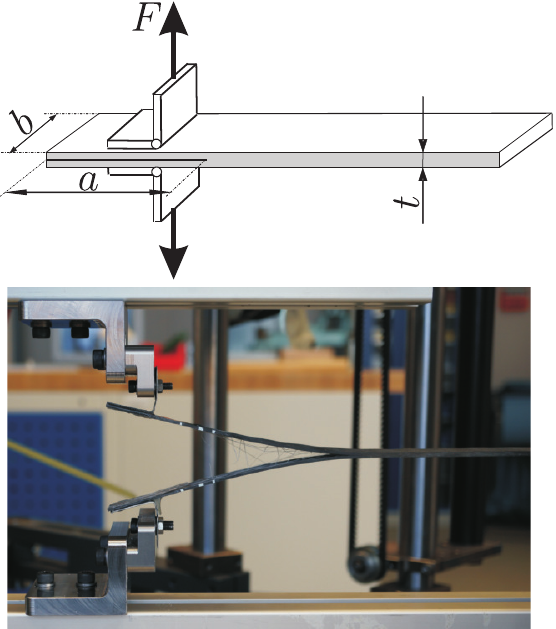}\qquad
    (b)~\includegraphics[scale=1.0]{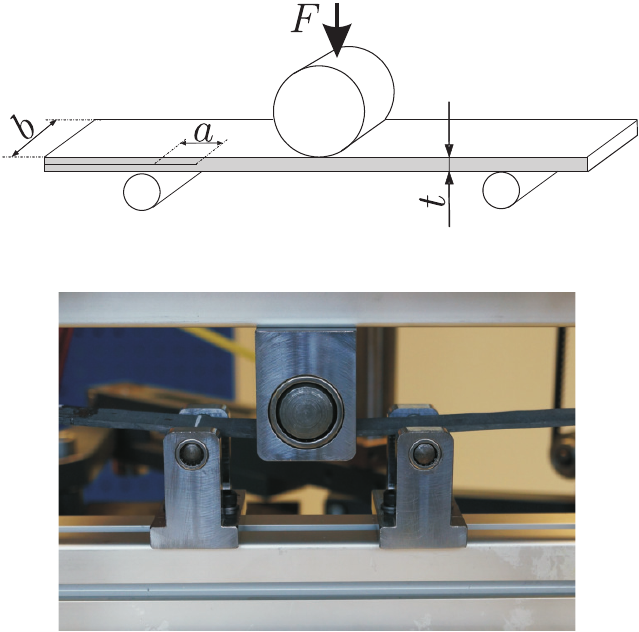}
    \caption{Experimental characterisation of the interface properties for intralaminar damage and failure: (a) DCB, and (b) ENF tests.}
  \label{fig:Exp_interface_damage_sketch}
  \end{center}
\end{figure}

To properly characterise the damage initiation and evolution for out-of-plane tensile loading, Double Cantilever Beam (DCB) tests, \reff{fig:Exp_interface_damage_sketch}~(a), have been carried out. The used specimens of width~$b=\SI{25}{mm}$ and thickness~$t=\SI{3}{\milli \metre}$ are made from unidirectionally reinforced laminates. Two stripes of Teflon have been inserted in between the middle layers during the lamination process in order to introduce an initial interface crack of length $a=\SI{25}{\milli\metre}$. The Mode I tensile loads are applied trough a hinge coupled to the testing machine.

Information regarding the in-plane-shear properties of the interface, i.e. Mode II loadings, can be obtained from End Notch Flexure (ENF) tests, \reff{fig:Exp_interface_damage_sketch}~(b). These experiments are essentially three point bending tests carried out on pre-notched specimens similar to the DCB test. The laminate has a thickness $t=\SI{3}{mm}$ and an initial crack length $a=\SI{35}{\milli\metre}$. 

\begin{figure}
  \begin{center}
    (a)~\includegraphics[]{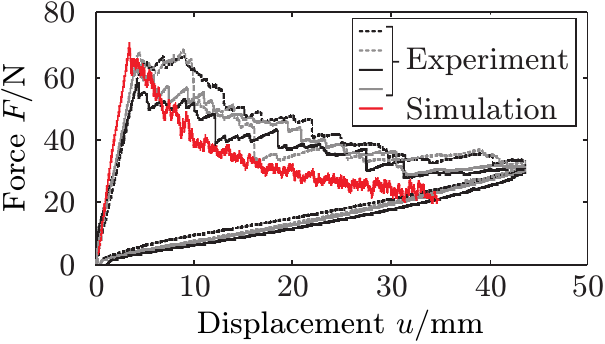}\qquad
    (b)~\includegraphics[]{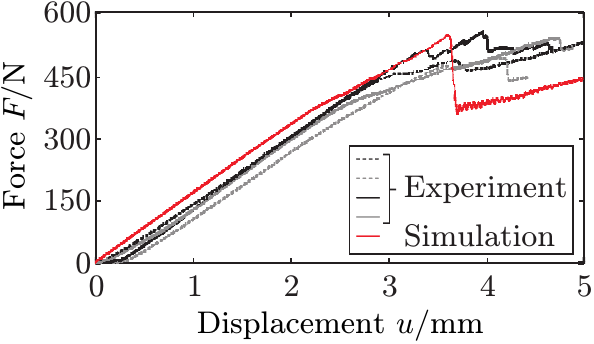}
    \caption{Comparison of simulated force-displacement curves to experimental results from (a) DCB, and (b) ENF tests.}
  \label{fig:Exp_interface_damage_diagram}
  \end{center}
\end{figure}

The displacement controlled DCB and ENF tests have been carried out based on DIN 6033/6034. The tests allow for the quantification of the critical energy release rates $G_{\text{Ic}}$ for Mode I, and $G_{\text{IIc}}$ for Mode II loadings, respectively. In combination with the obtained force-displacement curves presented in \reff{fig:Exp_interface_damage_diagram}, the parameters of a bi-linear traction-separation law according to \reff{fig:Methodical_Framework} have been determined. 

Assuming an isotropic shear behaviour, the parameters of the described cohesive zone model involve the maximum tractions $t^0_\text n$, and $t^0_\text s=t^0_\text t$ as well as the critical energy release rates $G_\text{Ic}=G_\text{nc}$, and $G_\text{IIc}=G_\text{sc}=G_\text{tc}$. These parameters have been chosen to provide a good agreement with the experimental force-displacement curves with the simulated mechanical response. The surface based cohesive model in ABAQUS is defined as an interaction property so that the elastic properties $k_i$ are based on the underlying element stiffness. In addition, the area under the traction-separation law must be equal to the corresponding critical energy release rate. The used parameters are summarised in \reft{tab:MPCohesive}. 
\begin{table}
\centering
\caption{Material parameters of the cohesive zone model as identified from DCB, and ENF tests.}
\begin{tabular}{p{2.7cm}cccc}
   \hline\hline
   Load Case& $G_{\text{c}}/\SI{}{\joule\per\square\meter}$ & $t^0/\SI{}{\newton\per\square\milli\meter}$\\\hline   
   Mode I & 543.6& 72.0 \\
   Mode II & 1214.8 & 72.0 \\\hline\hline
\end{tabular}
\label{tab:MPCohesive}
\end{table}

A comparison of the experimental and numerical results in \reff{fig:Exp_interface_damage_diagram} demonstrates that the stiffness of the specimen and the maximum force which marks the initiation of crack propagation are accurately recovered in the simulation for both tests. Nevertheless, the overall energy dissipated by crack propagation measured in terms of the hysteresis of the force-displacement curve is smaller than in the DCB test, \reff{fig:Exp_interface_damage_diagram}~(a). The simulation is also able to reproduce the characteristics of the ENF test where crack propagation leads to a steep decline of the applied force, \reff{fig:Exp_interface_damage_diagram}~(b). As the specimen is not fully cracked, a further increase of the displacement results in an increasing normal force. However, the stiffness of the specimen is reduced due the propagation of the interface crack.
\section{Simulation of a self-piercing rivetting process}
\label{sec:Simulation}
\begin{figure}
  \begin{center}
    \includegraphics[]{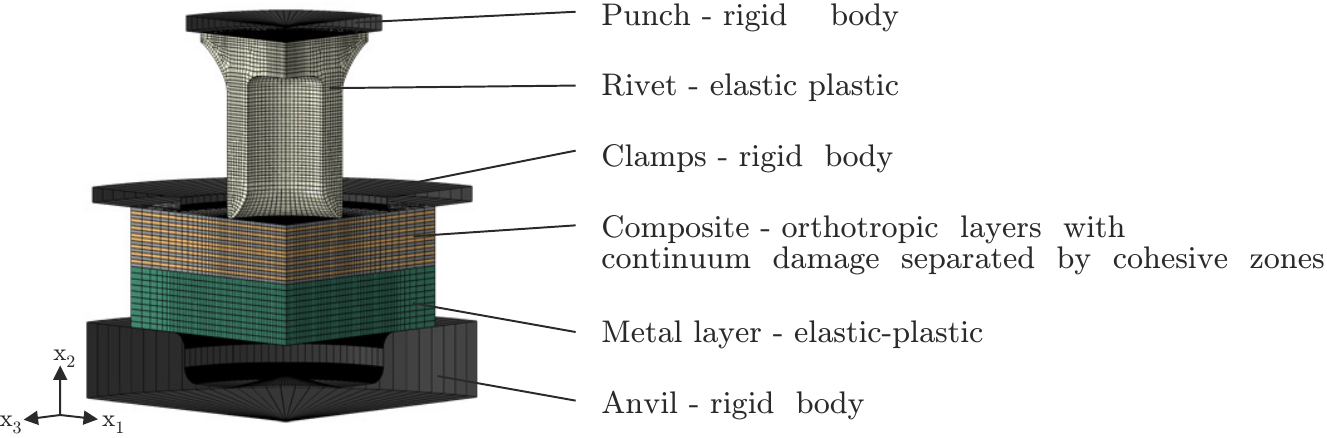}
    \caption{Model set-up for a process simulation of a composite/metal self-piercing rivetting process.}
  \label{fig:Process_simulation_set_up}
  \end{center}
\end{figure}

The material model for the composite material derived and parameterised in the previous sections is now applied in the numerical simulation of a self-piercing rivetting process. Figure \ref{fig:Process_simulation_set_up} illustrates the geometry of the individual parts and specifies the used material models. The model has been generated using the commercial FE software ABAQUS. Each deformable part is discretised by eight-node hexahedral elements with reduced integration and hourglass control (C3D8R). 

\subsection{Model set-up}

\begin{figure}
  \begin{center}
    (a)~\includegraphics[]{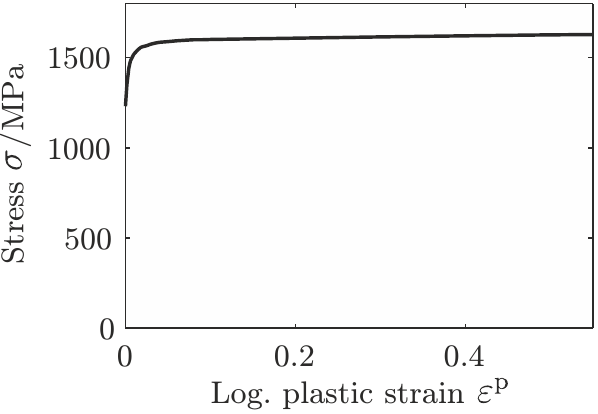}\qquad
    (b)~\includegraphics[]{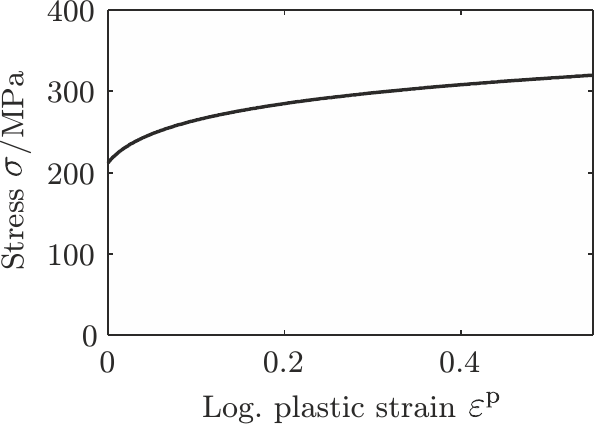}
    \caption{Flow curves determine the large deformation plastic behaviour in terms of the Cauchy stress $\sigma$ and the logarithmic plastic strain $\varepsilon^{\text{p}}$: (a) steel rivet, and (b) aluminium AW6060/T66 of the bottom sheet.}
  \label{fig:FlowCurves}
  \end{center}
\end{figure}

The deformation behaviour of the steel rivet and the aluminium bottom sheet is assumed to be elastic-plastic. The linear elastic behaviour up to the yield point is determined by the elastic constants $E_{\text{St}}=\SI{2.1e5}{\mega\pascal}$, $E_{\text{Al}}=\SI{6.9e4}{\mega\pascal}$, and $\nu_{\text{St}}=\nu_{\text{Al}}=\num{0.3}$. The flow curves illustrated in \reff{fig:FlowCurves} describe the large deformation plastic behaviour in terms of the Cauchy stress $\sigma$ and the logarithmic plastic strain $\varepsilon^{\text{p}}$.

\begin{figure}
  \begin{center}
    (a)~\includegraphics[]{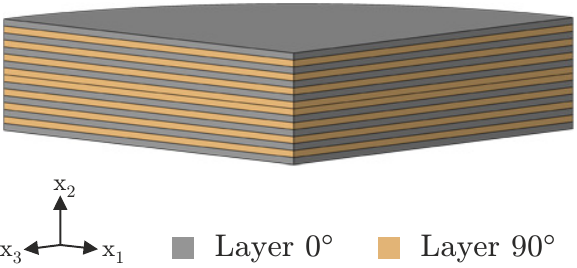}\qquad
    (b)~\includegraphics[]{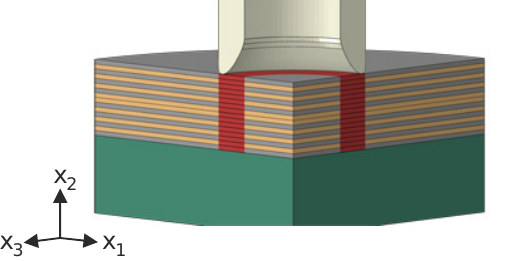}
    \caption{Details of the numerical process model: (a) biaxial composite lay-up, and (b) cutting zone with potential damage induced element erosion (red).}
  \label{fig:Process_simulation_set_up_damage}
  \end{center}
\end{figure}

The composite material consists of 16 individual layers with unidirectional reinforcement with a symmetric $\SI{0}{\degree}$/$\SI{90}{\degree}$ lay-up, cf. \reff{fig:Process_simulation_set_up_damage} (a). The process induced damage and failure of the composite material are captured by the orthotropic damage model of Section \ref{sec:Material} and the parameters identified in Section \ref{sec:Identification} in combination with the element erosion feature of ABAQUS. In order to prevent an excessive virtual material loss, the region for element erosion has been limited to the anticipated cutting zone, \reff{fig:Process_simulation_set_up_damage}~(b). Cohesive zone elements between the individual layers account for delamination using a bilinear traction-separation law and the parameters of Section \ref{sec:Cohesive}.

The modelling effort has been reduced substantially by exploiting the symmetry of the model, i.e. a quarter model has been used. While the geometry of the problem, and the loading would enable a two-dimensional axis-symmetric simulation, the actual lay-up of the composite only allows for the use of a quarter model. More complex stacking sequences could even require the analysis of a full three-dimensional model.

The total simulation time for the joining process has been set to \SI{0.005}{\second}, with a total displacement of \SI{6.05}{\milli\metre} for the punch. The load has been applied with a smoothed ramp function. The motion of the  workpieces has been constrained using symmetric boundary conditions at the external edges and by fixing the clamp position. The friction coefficient between the tools and the workpieces has been set to $\mu=0.30$, except for the contact of the rivet and the composite part, which is considered to be frictionless. The semi-automatic mass scaling feature of ABAQUS has been used to assure an average stable time increment of $\SI{1e-8}{\second}$. 

\subsection{Results and discussion}
\begin{figure}
  \begin{center}
    (a)~\includegraphics[]{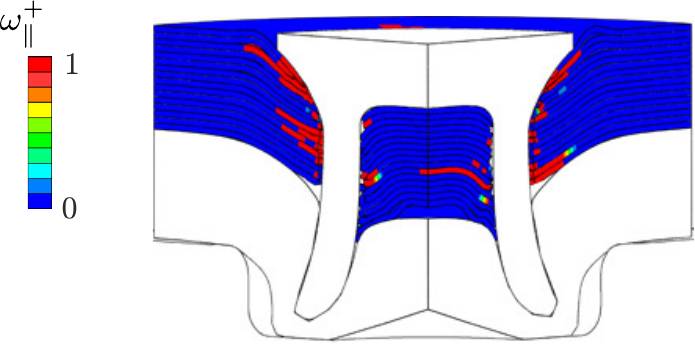}\qquad
    (b)~\includegraphics[]{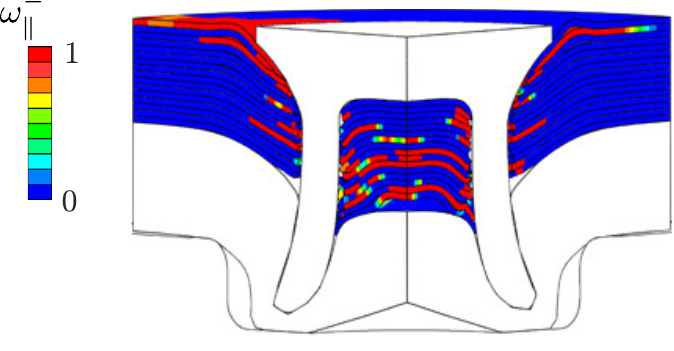}\\
    (c)~\includegraphics[]{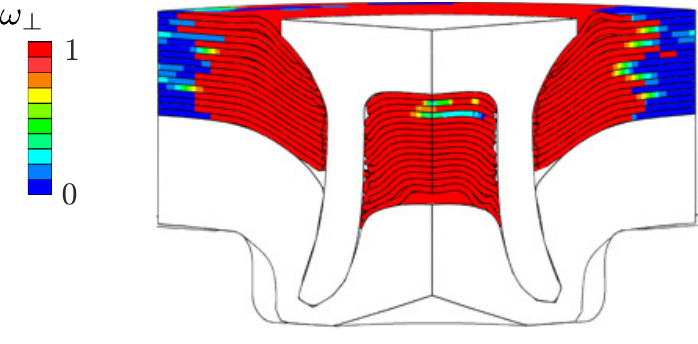}\qquad
    (d)~\includegraphics[]{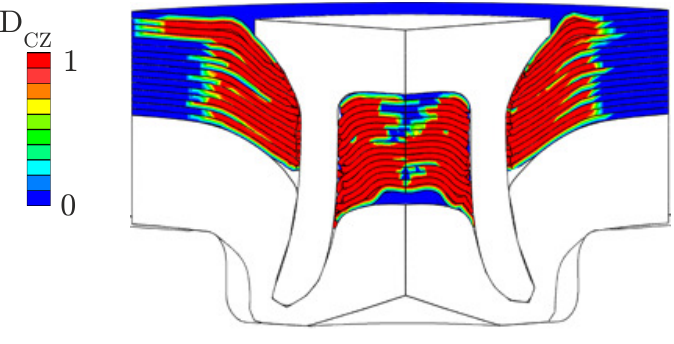}
    \caption{Process simulation of a composite/metal self-piercing rivet joint with a countersunk head rivet. Local distributions of: (a) fibre tension damage, (b) fibre compression damage, (c) matrix damage, and (d) delamination.}
  \label{fig:Process_simulation_countersunk_head}
  \end{center}
\end{figure}

It has been reported in various experimental investigations on SPR techniques for FRP/metal joints that the geometry of the rivet has a major influence on the SPR process and the quality of the joint. Therefore, two different rivets, with countersunk and flat heads, have been analysed numerically and experimentally. At first, the results for the countersunk head will be presented. The final configuration of the joint in combination with the local distributions of the damage thresholds for fibre and matrix failure as well as delamination are visualised in Figure \ref{fig:Process_simulation_countersunk_head}.

The obtained damage distribution essentially demonstrates the plausibility of the simulation approach and the used constitutive models. The bending deformation of the laminate produces in-plane tensile stresses that lead to fibre damage and failure close to the rivet. It is noted, that cut elements have been already removed from the model in \reff{fig:Process_simulation_countersunk_head}~(a). Compressive fibre failure is observed for laminae close to the surface of the
composite and the interior of the rivet, \reff{fig:Process_simulation_countersunk_head}~(b). The simulation results indicate that the matrix material is almost completely crushed in the vicinity of the rivet, in particular in zones directly below the rivet head, and in the interior of the rivet, cf. \reff{fig:Process_simulation_countersunk_head}~(c). Eventually, the delamination of the individual layers is illustrated in \reff{fig:Process_simulation_countersunk_head}~(d). It exhibits a distribution that is similar to the matrix failure mode.

\begin{figure}
  \begin{center}
    (a)~\includegraphics[]{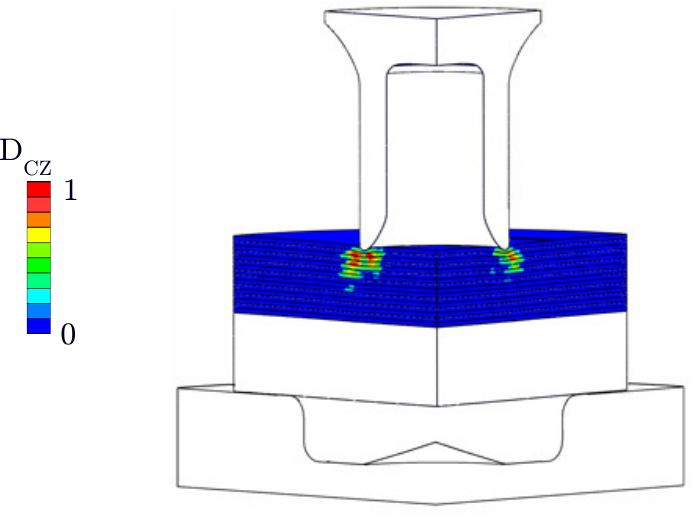}\qquad
    (b)~\includegraphics[]{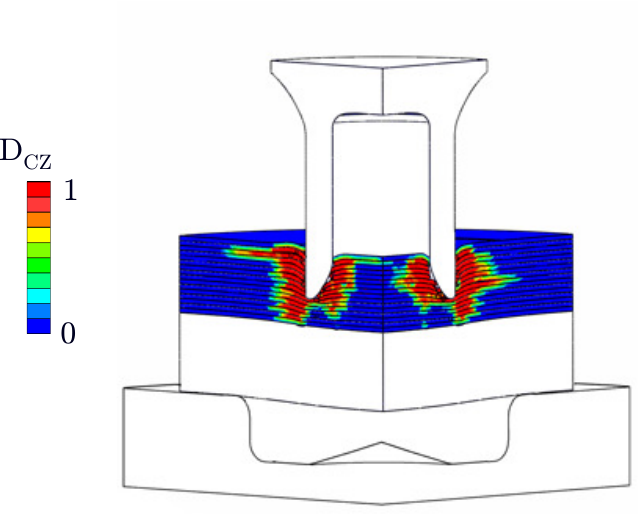}\\
    (c)~\includegraphics[]{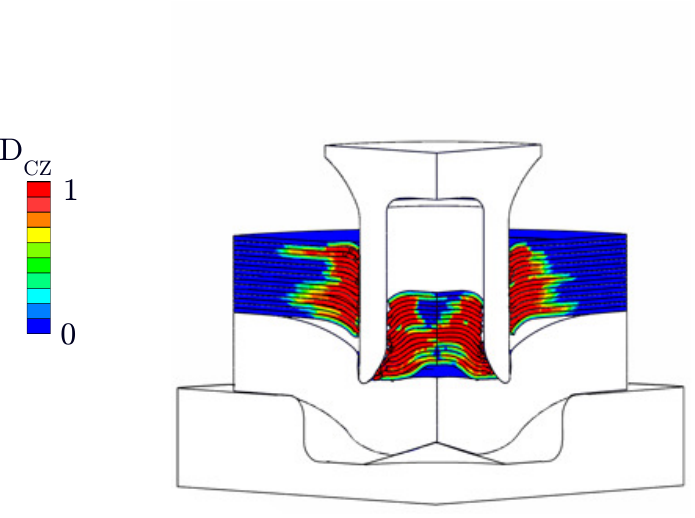}\qquad
    (d)~\includegraphics[]{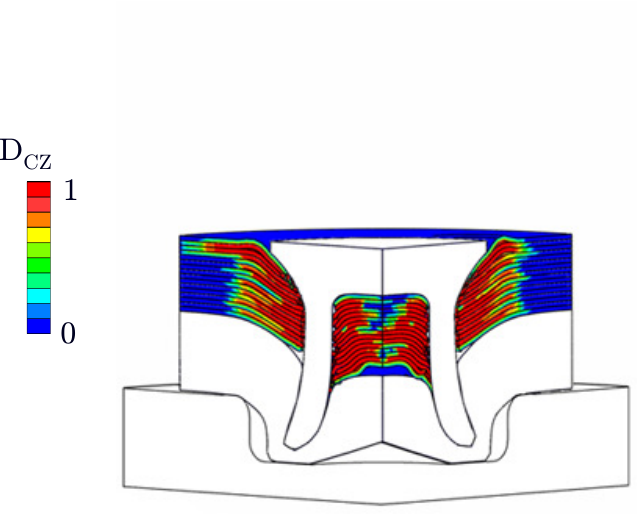}
    \caption{Distribution of interlaminar damage (delamination) at time $t$ during process of overall time $T$ of a composite/metal self-piercing rivet joint with a countersunk head rivet. (a) $t/T=0.2$, (b) $t/T=0.4$, (c)  $t/T=0.6$ and (d) $t/T=1$.}
  \label{fig:delamination_evolution_countersunk_head}
  \end{center}
\end{figure}

\label{page:delamination_evolution_countersunk_head} The evolution of interlaminar damage, i.e. delamination, during the piercing process is illustrated in \reff{fig:delamination_evolution_countersunk_head}~a) to d). The contour plots show the damage variable $D_\text{CZ}$ of the cohesive zone model at the normalized time $t/T$, with the actual simulation time $t$ and the overall process time $T$. Delamination initiates between the top layers due to the penetration of the rivet. With the advancing piercing process, the bending deformation of the composite leads to delamination of all laminae in the vicinity of the rivet. The intrusion of the rivet head at the end of the process in \reff{fig:delamination_evolution_countersunk_head}~d) significantly extends the interlaminar failure in the radial direction.

\begin{figure}
  \begin{center}
    (a)~\includegraphics[]{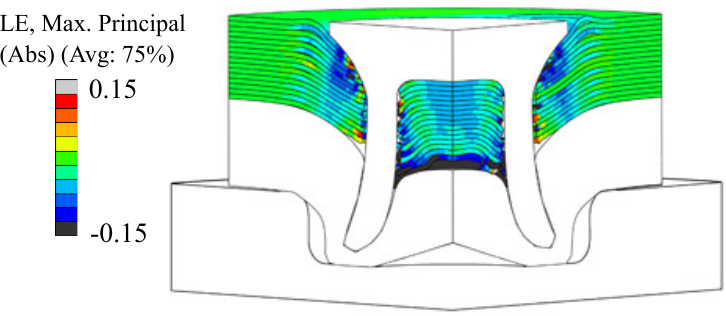}\qquad
    (b)~\includegraphics[]{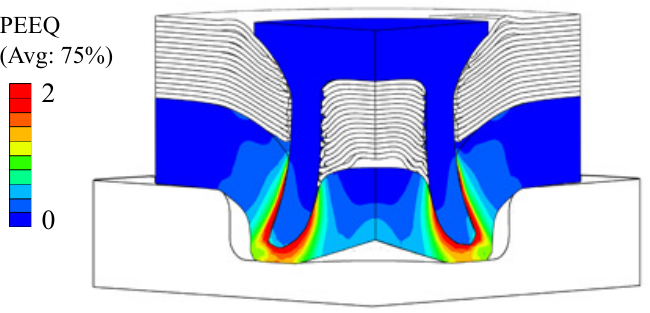}
    \caption{Distribution of local strains in the composite and the metal sheet: (a) Maximum principal logarithmic strain in the composite, and (b) equivalent plastic strain in the aluminium sheet.}
  \label{fig:strain_distribution_countersunk_head}
  \end{center}
\end{figure}

\begin{figure}
  \begin{center}
    (a)~\raisebox{1.6cm}[0pt]{\includegraphics[width=0.35\textwidth]{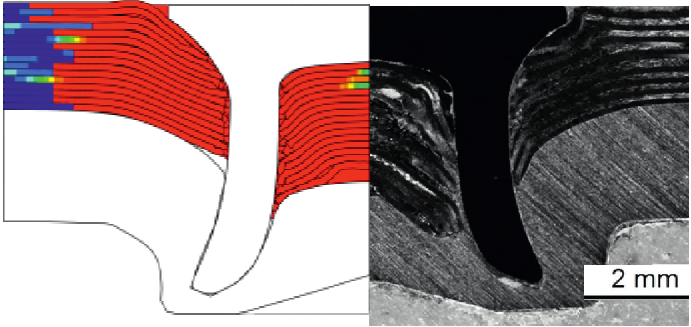}}
    (b)~\includegraphics[width=0.5\textwidth]{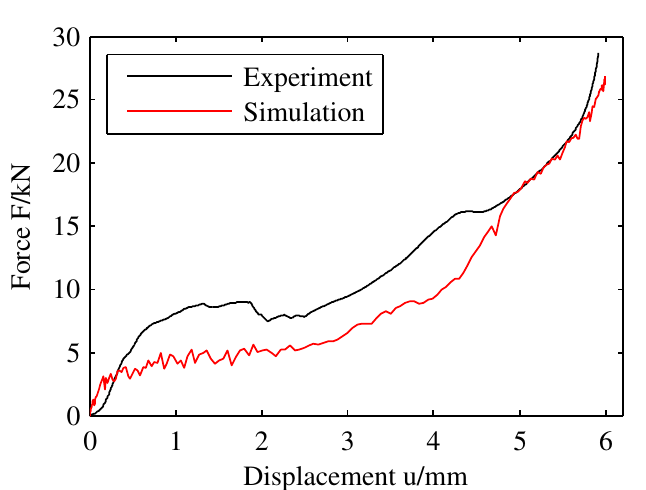}
    \caption{Comparison of numerical and experimental results obtained for the countersunk head rivet: (a) deformation in the joining zone with colours indicating damage, and (b) force-displacement curves of the punch.}
  \label{fig:SimCountersunk}
  \end{center}
\end{figure}

\label{page:strain_distribution_countersunk_head} As outlined in Section \ref{sec:Material}, the applied composite damage model is based on a small strain theory and accounts for large rotations in terms of a corotational formulation. To confirm the constitutive assumptions and to assess the occurring strain levels, the maximum principal strain within the composite material and the equivalent plastic strain within the metal components are shown in \reff{fig:strain_distribution_countersunk_head}~a) and b), respectively. It is found from \reff{fig:strain_distribution_countersunk_head}~a) that the principal strains are less than $10\%$. As expected, the small strain range is exceeded only in small areas close to the rivet head and in the bottom layer. However, the composite in these domains is fully disintegrated due to damage and the element stiffness represents rather a numerical value than a calibrated material property. For applications, where a composite undergoes large deformations, e.g. during the forming process of metal-composite clinching joints, a composite model based on finite deformation theory will be required. \cite{dean2016} recently proposed a model for short fiber reinforced polymers. In contrast, the aluminium sheet undergoes large plastic deformation to form into the rigid anvil, \reff{fig:strain_distribution_countersunk_head}~b).
\newpage
The comparison of the simulated final geometry of the joining zone to a micrograph image of the corresponding joining experiment in \reff{fig:SimCountersunk}~(a) shows that the simulation is able to represent the principal characteristics of the joining zone. However, it cannot reproduce the severity of the material damage in particular in the interior of the rivet. Moreover, the deflection of the composite layers close to the bottom sheet is underestimated. The reason for these discrepancies is assumed to be the virtual volume loss due to element erosion which leads to a significant reduction of friction effects and additional deformation that result from the destroyed material still being present in the joining zone. \label{page:volume_loss} The virtual volume loss can amount up to $8.2\%$ with respect to the composite layer or $3.5\%$ with respect to whole model.

That is, the deformation of the composite and the resulting damage in the vicinity of the rivet is in fact mainly caused by contact to the tapered rivet head. During the actual piercing process elements directly below the rivet quickly reach a damage level that causes the immediate deletion of the affected element. This assumption is backed by the comparison of the numerical and experimental force-displacement curves in \reff{fig:SimCountersunk}~(b). As a consequence of element erosion, the simulation underestimates the required force level during the piercing of the composite. This could also be attributed to the general underestimation of the intralaminar shear strength in the experimental characterisation procedure, since the primary mechanism of deformation in the experiments appears to be delamination induced by shearing, cf. \reff{fig:SimCountersunk}~a). \label{page:shear}

\begin{figure}
  \begin{center}
    (a)~\raisebox{1.6cm}[0pt]{\includegraphics[width=0.35\textwidth]{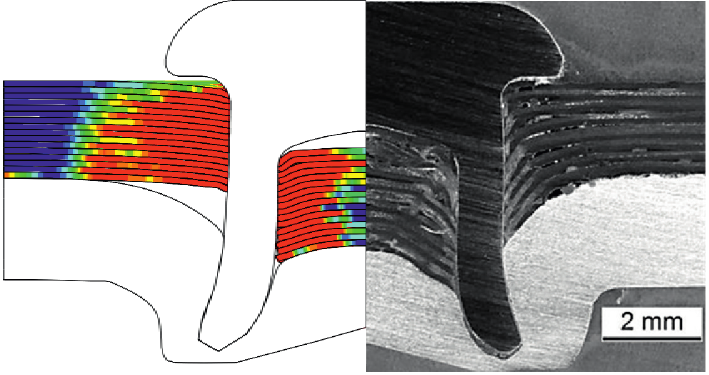}}
    (b)~\includegraphics[width=0.5\textwidth]{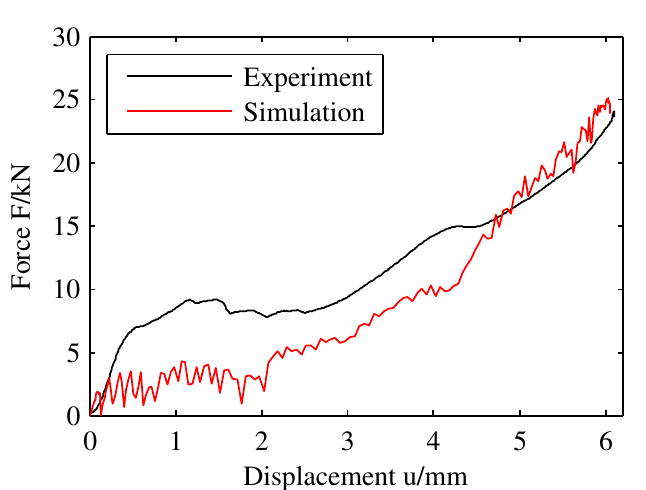}
   \caption{Comparison of numerical and experimental results obtained for the flat head rivet: (a) deformation in the joining zone with colours indicating damage, and (b) force-displacement curves of the punch.}
  \label{fig:SimFlat}
  \end{center}
\end{figure}

The effect of element erosion on the deformation of the composite and the resulting damage pattern is even more dramatic in the case of the flat head rivet, \reff{fig:SimFlat}. As there is no direct contact of the head to the laminate, virtually no deflection of the composite can be induced. Consequently, there is much less intra- and interlaminar damage in the zone outside the rivet. Failure localises in elements directly below the rivet which are removed during the simulation process. These problems translate directly into a force-displacement curve that clearly underestimates the force that is required to pierce the laminate.

\section{Conclusions and perspectives}
\label{sec:Conclusion}

In this contribution, a numerical framework for the simulation of self-piercing rivetting processes in FRP laminates and sheet metals has been presented. Special emphasis was put on the modelling of the deformation and failure behaviour of the composite which is governed by intra- and interlaminar damage phenomena. Depending on the used polymeric matrix, rate dependent viscoelastic effects can interfere with the strength and the long-term behaviour of the joint. A systematic approach to the characterisation and the modelling of composite materials has been outlined and the limitations of the model and the parameter identification were discussed. 

The macroscopic, orthotropic, potentially viscoelastic mechanical properties of the composite have been examined using a two-scale homogenisation technique. This method requires a geometric model of the local material structure and the mechanical behaviour of the individual constituents but provides insight in structure-property relations of the composite and allows the prediction of the effective behaviour for different material combinations and reinforcements. 
To analyse the local material damage caused by the piercing process, an orthotropic continuum damage model has been parameterised based on tension and compression tests on unidirectionally reinforced composites. In order to represent delamination, a bi-linear cohesive zone model has been parameterised based on DCB and ENF tests.

The proposed modelling framework has eventually been applied to the simulation of a self-piercing riveting process. The evaluation of a flat and a countersunk head rivet shows a substantial dependence of the local composite damage distribution on the rivet geometry. Different from previously applied modelling approaches, the presented simulation results have to be considered a prediction as a clear differentiation of intra- and interlaminar damage processes and a systematic parameter identification have been used. However, the virtual material loss caused by the element erosion leads to an underestimation of the joining force, especially during the piercing of the composite laminate. Future work will therefore have to address the issue of virtual material loss in terms of an improved parameterisation of the damage model and its interaction with element erosion. Alternatively, different strategies to model the failure of the individual layers should be evaluated. E.g. the local breakage of node-ties or non-local formulation of the continuum damage could help to avoid the excessive material loss but they are currently not available in commercial FE programmes. 

\section*{Acknowledgement}
The present study is supported by the German Federation of Industrial Research Associations (AiF) in collaboration with the European Research Association for Sheet Metal Working (EFB) within the project 17594BR (AiF) / 06-211 (EFB). This support is gratefully acknowledged.
\\

\bibliographystyle{elsarticle-harv}
\bibliography{JoMP2016}

\end{document}